\documentclass[apj]{emulateapj}
\usepackage{apjfonts}

\slugcomment{Accepted for publication in The Astrophysical Journal}
\shorttitle{SHEAR OF SN Ia PROGENITORS}
\shortauthors{PIRO}

\newcommand{\be}{\begin{eqnarray}}
\newcommand{\ee}{\end{eqnarray}}
\newcommand{\lp}{\left(}
\newcommand{\rp}{\right)}
\newcommand{\wk}{\Omega_{\rm K}}
\newcommand{\tvisc}{t_{\rm visc}}
\newcommand{\tacc}{t_{\rm acc}}

\begin{document}


\title{The Internal Shear of Type Ia Supernova Progenitors During Accretion and Simmering}

\author{Anthony L. Piro}

\affil{Astronomy Department and Theoretical Astrophysics Center,
601 Campbell Hall,\\ University of California,
Berkeley, CA 94720; tpiro@astro.berkeley.edu}


\begin{abstract}
   A white dwarf (WD) gains substantial angular momentum during the
accretion process that grows it toward a Chandrasekhar mass. It is therefore
expected to be quickly rotating when it ignites as a Type Ia supernova.
The thermal and shearing profile are important for subsequent flame propagation.
We highlight processes that could affect the WD shear, during accretion as
well as during the $\sim1000$ years of pre-explosive simmering.
Baroclinic instabilities and/or the shear growth of small magnetic fields
provide sufficient torque to bring the WD very close to solid body rotation during accretion.
The lack of significant shear makes it difficult to grow a WD substantially past the
typical Chandrasekhar mass. Once carbon ignites, a convective region spreads from the
WD's center. This phase occurs regardless of
progenitor scenario, and therefore it is of great interest for understanding
how the WD interior is prepared before the explosive burning begins. We summarize some of the key
properties of the convective region, which includes demonstrating that the mass enclosed
by convection at any given time
depends most sensitively on a single parameter that can be expressed as either
the ratio of temperatures or densities at the top and bottom
of the convection zone. At low Rossby numbers the redistribution of angular
momentum by convection may result in significant shearing at the convective/non-convective
boundary.
\end{abstract}



\section{Introduction}

   The fundamental role Type Ia supernovae (SNe Ia) play in determining the
expansion history of the universe \citep[][and references therein]{rie04} has
added new urgency to understanding the progenitors of these events.
Theoretical work on modeling the flame dynamics of exploding white dwarfs (WDs) has
demonstrated that the composition and energy of ejecta depend sensitively
on the competition between flame propagation, instabilities driven by
turbulence, and expansion of the WD \citep[e.g.,][]{hn00}. This has sparked new interest in
understanding the evolution that occurs up until the onset of explosive burning.
A critical aspect is the shear profile present within the
WD.

   SNe Ia are expected to occur exclusively in accreting binary systems. 
Therefore the role of angular momentum may be unique or at least especially
important for these SNe. If accretion drives shear in the WD, it may be
an important source of viscous heating and material mixing.
\citet{yl04} considered the accretion and evolution
of a WD up until the point of ignition, including the effects of rotation.
In their study angular momentum was transported primarily via the
Kelvin-Helmholtz instability, which lead to significant shear throughout the WD.
\citet{sn04} focused on whether rotational effects can
prevent accretion induced collapse (AIC) at high accretion rates
($\dot{M}=3\times10^{-6}-10^{-5}\ M_\odot\ {\rm yr^{-1}}$). They found that an AIC
was not averted,
but perhaps more interestingly, they also found the WD should
be rotating nearly uniformly when the baroclinic instability is included, a
process that was neglected by \citet{yl04}.

   There is further opportunity for shear to develop during
the carbon simmering phase. At carbon ignition, a convective region grows at
the WD center. This envelops $\sim1\ M_\odot$ over a timescale
of  $\sim1000\ {\rm yrs}$, until a burning wave commences.
This phase has received increasing attention in recent years
due to the realization that it sets the initial temperature and density for the explosive
burning \citep{les06}, as well as the distribution of ignition points
\citep{woo04,ww04,kuh06}. Nuclear reactions during this time
may change the neutron excess of the WD core, which is crucial for
determining the ratio of radioactive to non-radioactive nickel produced in the Type Ia
explosion \citep{pb07b,cha07}.

    In the present study we consider the opportunity for developing shear during
both of these stages. In \S \ref{sec:accretion} we revisit
estimates for the shear of accreting WDs and confirm that the baroclinic instability
limits the shear before the Kelvin-Helmholtz instability initiates. In addition we explore whether
magnetohydrodynamic effects could reduce the shear even further. We conclude
that the WD is nearly uniformly rotating at the onset of
carbon ignition, and that viscous heating is negligible. In \S \ref{sec:convection} we
explore the properties of convection present prior to explosive burning.
The growth of this convective region allows further opportunity for shearing.
The three-dimensional nature of this
problem makes it hard to definitively determine what occurs during this stage.
We illustrate some general features expected for the interaction of spin and convection
by appealing to observations and numerical experiments.
In \S \ref{sec:theend} we conclude with a summary of our results and a discussion
of future research.


\section{Shear Profile During Accretion}
\label{sec:accretion}

   Material accreted at a rate $\dot{M}$ reaches the WD surface with a nearly Keplerian
spin frequency of $\wk=(GM/R^3)^{1/2}$. The majority of the kinetic energy associated
with this flow is dissipated in a boundary layer of thickness $H_{\rm BL}\ll R$
\citep[as studied by][]{pb04} and never reaches far into the surface. Nevertheless,
angular momentum is added at a rate $\dot{M}R^2\wk$, so a
torque of this magnitude should be communicated into the WD. It remains an open question
whether all of this angular momentum is ultimately added to the WD. If it was, the WD would
reach overcritical rotation well before attaining the conditions necessary for
carbon ignition \citep{mac79}. Feedback with the accretion disk
at high spin rates may provide a solution to this problem \citep{pac91,pn91}.
Observations of cataclysmic
variables show a wide range of rotational velocities, many of which are too low
in comparison to the amount of angular momentum that should have been accreted
\citep{sio99}. This may be related to the periodic mass loss these systems
undergo in classical novae \citep{lp98}.

   For our present study the viscous mechanisms we consider have viscous timescales,
 $\tvisc=H^2/\nu$, where $H=P/\rho g$ is the pressure scale height, $g=GM/r^2$ is the local
gravitational acceleration, and $\nu$ is the viscosity\footnote{This viscosity can represent
either a molecular viscosity or a turbulent viscosity, but for the scenarios discussed in \S 2.1
a turbulent viscosity is argued to be dominant.}, that are much shorter
than the timescale of accretion, $\tacc\sim10^5-10^9\ {\rm yrs}$. Due to this hierarchy
of timescales we expect that the spin at any given moment can be broken into
two contributions, which we write as
\be
	\Omega(r,t) = \Omega_0(t)+\Delta\Omega(r,\Omega_0(t)).
	\label{eq:omega}
\ee
The first piece, $\Omega_0(t)$, represents the solid-body rotation of the WD, which is
increasing with time as accretion takes place. The second piece, $\Delta\Omega(r,\Omega_0(t))$,
is the shear that must be present to transport the accretion torque into the WD.
As long as the timescales obey $\tvisc/\tacc\ll1$, the viscosity is extremely efficient at
transporting angular momentum. This means that the shear needed for transport is
small, $\Delta\Omega\ll\Omega_0$, and that the shear quickly comes into steady-state
for a given $\Omega_0(t)$, so that $\Delta\Omega$ does not explicitly depend on $t$.

   Since all of the transport
mechanisms we examine operate most efficiently in directions perpendicular to gravity
(because no work is performed), it is adequate to consider
a structure composed of concentric spheres, each with constant $\Omega$.
Transfer of angular momentum
is reduced to a one-dimensional diffusion equation \citep{fuj93}
\be
	\frac{\partial}{\partial t}(r^2\Omega) = \frac{1}{r^2\rho}\frac{\partial}{\partial r}
		\lp\rho \nu r^4 \frac{\partial\Omega}{\partial r}\rp.
\ee
We substitute equation (\ref{eq:omega}) and assume that $r$ is roughly independent
of time to obtain
\be
	r^2\lp1+\frac{\partial\Delta\Omega}{\partial\Omega_0}\rp\frac{d{\Omega}_0}{dt}
	=\frac{1}{r^2\rho}\frac{\partial}{\partial r}
		\lp\rho \nu r^4 \frac{\partial\Delta\Omega}{\partial r}\rp.
\ee
The second term within the parenthesis on the left hand side is negligible
since it is $O(\Delta\Omega/\Omega_0)\ll1$.
Multiplying both sides by $4\pi\rho r^2$ and integrating over radius we
find
\be
	I(r)\frac{d\Omega_0}{dt} = 4\pi \rho \nu r^4 \frac{\partial\Delta\Omega}{\partial r},
\ee
where we have set the integration constant to
zero to assure that the torque vanishes at the WD center and
\be
	I(r) = \int_0^r 4\pi \rho r^4dr
\ee
is the moment of inertia interior to a given radius. The rate of change of the
WD spin is set by the accretion torque, so that
$I_{\rm tot}d\Omega_0/dt=\dot{M}R^2\wk$, where $I_{\rm tot}=I(R)$ is the
total moment of inertia. This gives us an equation for the shear
$\sigma\equiv\partial\Delta\Omega/d\ln r$,
\be
	\dot{M}R^2\Omega_{\rm K}I(r)/I_{\rm tot}= 
	4\pi\rho \nu r^3 \sigma.
	\label{eq:angularmomentum}
\ee
This result reduces to the angular momentum equations presented by \citet{fuj93} and
\citet{pb07a} in the plane parallel limit by taking $I(r)\approx I_{\rm tot}$.
With this expression we can now estimate the shear at a given depth as a function
of $\Omega_0$ and $\dot{M}$.  The advantage of this approach is
that $\Omega_0$ can be treated as a free parameter, and we are not restricted
to follow the full history of angular momentum transport for the WD.

   Since for the remainder of \S \ref{sec:accretion} we assume that the approximations
presented here hold true, we simplify notation
by writing $\Omega$ as the solid-body rotation rate and $\sigma$
as the shear.

\subsection{Summary of Transport Mechanisms}

   In the following sections we discuss some of hydrodynamic and magnetohydrodynamic
instabilities expected to be present in the shearing WD core. We summarize three distinct eddy
diffusivities related to these instabilities and explore their influence on the shear profile.
This is not meant
to be an exhaustive survey of all turbulent angular momentum transport
mechanisms possible. Instead it is meant to
highlight those that have been used in previous studies along with others
that have received less attention in the past, but which we show to be important.

\subsubsection{Kelvin-Helmholtz Instability}
\label{sec:kh}

   A hydrodynamic instability that has been popular for application to accreting
WDs is the Kelvin-Helmholtz instability \citep[see][and associated work]{yl04}.
This instability is activated when $Ri<1/4$, where
the Richardson number is
\be
	Ri \equiv \frac{N^2}{\sigma^2},
	\label{eq:richardson}
\ee
and $N$ is the Brunt-V\"{a}is\"{a}l\"{a} frequency, given by
\be
	N^2 = \frac{\mathcal{Q}g}{H}
		\left[\nabla_{\rm ad}-\lp\frac{d\ln T}{d\ln P}\rp_*\right],
\ee
(ignoring compositional gradients) where $\mathcal{Q}=-(\partial\ln \rho/\partial\ln T)_P$,
$\nabla_{\rm ad}=(\partial\ln T/\partial\ln P)_{\rm ad}$
is the adiabatic temperature gradient, and the asterisk refers to derivatives of the envelope's
profile. The electron viscosity is the dominant molecular viscosity in the degenerate WD.
Substituting the viscosity from \citet{np84}
into equation (\ref{eq:angularmomentum}) results in
a shear in which $Ri\ll1/4$, and thus Kelvin-Helmholtz instability
is expected. This confirms
the result of \citet{yl04} that in the absence of other instabilities, Kelvin-Helmholtz instability
is dominant, so that we view its associated shear as an upper limit.

   Kelvin-Helmholtz instability causes a turbulent eddy diffusivity given by \citep{fuj93}
\be
	\nu_{\rm KH}=\frac{(1-4Ri)^{1/2}}{2Ri^{1/2}}H^2N.
\ee   
Since it only acts for $Ri<1/4$, it causes the shearing to
evolve until $Ri=1/4$ is marginally satisfied. Thus it is a good
approximation to estimate the shear to be
$\sigma_{\rm KH}\approx2N$ when secular instabilities due to thermal
diffusion can be neglected \citep[e.g.,][]{zah92}, as is the case for the conductive WD core.
Since electron degeneracy provides the dominant pressure, we approximate
the Brunt-V\"{a}is\"{a}l\"{a} frequency as
\be
	N\approx\lp\frac{g}{H}\frac{k_{\rm B}T}{ZE_{\rm F}}\rp^{1/2},
	\label{eq:brunt}
\ee
where $k_{\rm B}$ is Boltzmann's constant, $Z$ is the charge per ion, and
$E_{\rm F}$ is the Fermi energy
for a degenerate, relativistic electron gas. We estimate for Kelvin-Helmholtz
driven shear,
\be
	\sigma_{\rm KH} \approx 0.5\ {\rm s^{-1}}\ g_{10}
		\lp\frac{\mu_e}{2}\rp^{5/6}\lp\frac{6}{Z}\rp^{1/2}\frac{T_8^{1/2}}{\rho_9^{1/3}},
	\label{eq:sigmakh}
\ee
where $g_{10}=g/10^{10}\ {\rm cm\ s^{-2}}$, $\mu_e$ is the mean molecular weight per
electron, $\rho_9=\rho/10^9\ {\rm g\ cm^{-3}}$, and $T_8=T/10^8\ {\rm K}$. This result is in
reasonable agreement to the detailed, time dependent simulations by \citet{yl04}.

\subsubsection{Baroclinic Instability}
\label{sec:bc}

   Another hydrodynamic instability that may be important is
the baroclinic instability (Fujimoto 1987, 1988; also see Cumming \& Bildsten 2000).
This instability arises because surfaces of constant pressure and density no longer
coincide when hydrostatic balance is maintained under differential rotation.
In such a configuration, fluid perturbations along nearly horizontal directions are
unstable, though with a sufficient radial component to allow mixing of angular
momentum and material. When $Ri$ is greater than the critical baroclinic Richardson
number \citep{fuj87},
\be
	Ri_{\rm BC} \equiv 4\lp\frac{r}{H}\rp^2\lp\frac{\Omega}{N}\rp^2,
\ee
Coriolis effects limit the horizontal scale of
perturbations and the associated turbulent viscosity is approximated from
linear theory to be \citep{fuj93}
\be
       \nu_{\rm BC} =  \frac{1}{3}\frac{Ri_{\rm BC}}{Ri^{3/2}}H^2\Omega.
       \label{eq:nubc}
\ee
In the WD interior $Ri_{\rm BC}\sim10$, whereas the Richardson number found
for $\nu=\nu_{\rm BC}$ (using eqs. [\ref{eq:angularmomentum}] and [\ref{eq:richardson}])
is $Ri\sim10^6$, thus the limit $Ri\gg Ri_{\rm BC}$
is clearly satisfied.

   The Richardson number can be estimated by combining equations
 (\ref{eq:angularmomentum}), (\ref{eq:richardson}), and
 (\ref{eq:nubc}), giving 
\be
	Ri^2=\frac{16\pi}{3}\frac{\rho r^3\Omega}{\dot{M}}
		\lp\frac{r}{R}\rp^{2}\frac{\Omega}{\wk}
		\frac{\Omega}{N}\frac{I_{\rm tot}}{I(r)}.
\ee
To illustrate how the shear depends on the properties of the interior we
take
\be
	\frac{16\pi}{3}\frac{\rho r^3\Omega}{\dot{M}}
		\sim\frac{\tacc}{t_{\rm dyn}} \frac{\Omega}{\wk}\lp\frac{r}{R}\rp^3,
\ee
where $\tacc=M/\dot{M}$ is the accretion timescale and $t_{\rm dyn}=\wk^{-1}$ is the
dynamical time. Using $\sigma_{\rm BC}=N/Ri^{1/2}$ and $I(r)/I_{\rm tot}\sim(r/R)^5$, we
estimate
\be
	\sigma_{\rm BC} &\sim&10^{-4}\ {\rm s^{-1}}
		\ \lp\frac{\tacc/t_{\rm dyn}}{10^{15}}\rp^{-1/4}
		\lp\frac{\Omega/\wk}{0.1}\rp^{-1/2}
		\nonumber
		\\
		&&\times
		\lp\frac{N}{\Omega}\rp^{1/4} \lp\frac{N}{0.3\ {\rm s^{-1}}}\rp.
	\label{eq:sigmabc}
\ee
In \S \ref{sec:timescale} we show that the timescale for viscous diffusion from
the baroclinic instability is much shorter than the timescale over which accretion
takes place. This means that this instability is able to quickly redistribute
angular momentum as it accretes and the steady-state limit will be reached.
Coupling this fact to the inequality
$\sigma_{\rm BC}\approx\sigma_{\rm KH}/(2Ri)\ll\sigma_{\rm KH}$ as
demonstrated above,
we conclude that the baroclinic instability will limit the growth of shear
long before the Kelvin-Helmholtz
instability can become active. This is consistent with the results of
\citet{sn04} who included both the Kelvin-Helmholtz instability and the baroclinic
instability and found nearly solid body rotation throughout the WD, even for
accretion rates as high as $10^{-5}\ {M_\odot\rm\ yr^{-1}}$.

\subsubsection{Tayler-Spruit Dynamo}
\label{sec:ts}

   The last case we consider is magnetohydrodynamic instabilities, for which
we apply the Tayler-Spruit dynamo \citep{spr02}.
In this picture, shearing stretches any small component of poloidal magnetic field
into a strong toroidal field. Once sufficiently large, the toroidal field
initiates Tayler instabilities \citep[non-axisymmetric, pinch-like instabilities
including stratification,][]{tay73,spr99}, which turbulently create poloidal field
components that once again shear to be toroidal. This cycle continues and results in a
steady-state field that transmits angular momentum via Maxwell stresses.
In the limit when the magnetic diffusivity, $\eta$, is much less
than the thermal diffusivity, $K$, the minimum shear needed for this
process to activate is \citep{spr02},
\be
	\sigma_{\rm TS,crit} = \lp\frac{N}{\Omega}\rp^{7/4}\lp\frac{\eta}{r^2N}\rp^{1/4}
		\lp\frac{\eta}{K}\rp^{3/4}\Omega.
	\label{eq:sigmatscrit}
\ee
When $\sigma>\sigma_{\rm TS,crit}$, the effective viscosity due to the
steady-state magnetic fields is
\be
	\nu_{\rm TS}=r^2\Omega\lp\frac{\Omega}{N}\rp^{1/2}
				\lp\frac{K}{r^2N}\rp^{1/2}.
\ee
We wait until the
following section to present the shear associated with this process.

\subsection{Comparison Calculations}
\label{sec:comparison}

   We now calculate isothermal WD profiles and the shear profiles
for the mechanisms described above. Such models allow us to
argue that {\it solid body rotation} is the most likely result.

  The WD models are computed by solving
for hydrostatic balance with the effects of spin ignored since we generally
consider spins of the order of $0.1\Omega_{\rm K}$.
We solve for $\rho$ using the analytic equation
of state from \citet{pac83}.
The importance of Coulomb interactions is measured
by the parameter
\be
	\Gamma=\frac{(Ze)^2}{ak_{\rm B}T}
		=  35.7\frac{\rho_9^{1/3}}{T_8}\lp\frac{Z}{6}\rp^2\lp\frac{12}{A}\rp^{1/3},
\ee
where $A$ is the mass per ion and $a$ is the ion separation.
For the liquid phase, when $1\le\Gamma\le173$,
we include the ionic free energy of \citet{cp98}.
The composition is set to 50\% $^{12}$C, 48\% $^{16}$O, and 2\% $^{22}$Ne
by mass.

\begin{figure}
\epsscale{1.2} 
\plotone{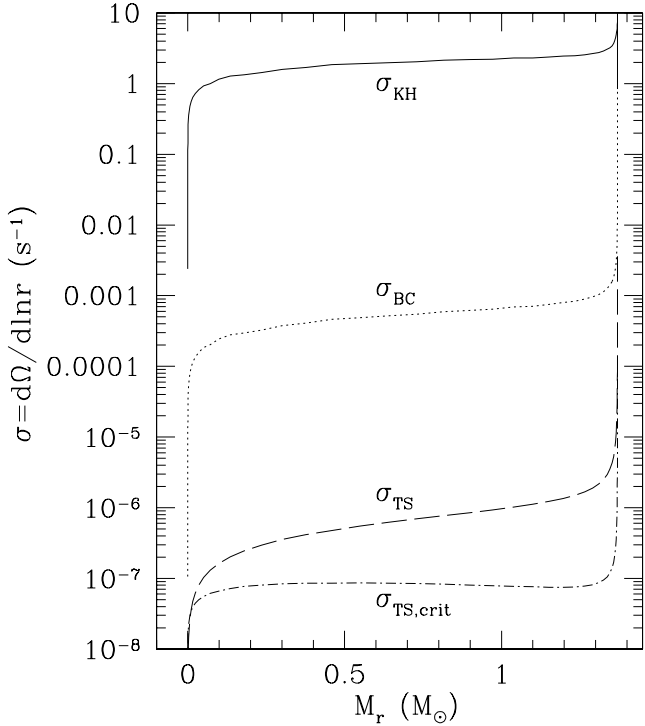}
\caption{Comparison of the shear rate associated with three different
viscous mechanisms as a function of mass coordinate:
Kelvin-Helmholtz instability ({\it solid line}),
baroclinic instability ({\it dotted line}), and the Tayler-Spruit dynamo
({\it dashed line}). We also plot the critical shear rate needed
for the Tayler-Spruit dynamo to be activated ({\it dot-dashed line}).
The WD is isothermal, with $T_i=10^8\ {\rm K}$,
$M=1.37\ M_\odot$, $\Omega=0.1\wk=0.67\ {\rm s^{-1}}$,
and $\dot{M}=10^{-7}\ M_\odot\ {\rm yr}^{-1}$.
}
\label{fig:comparison}
\epsscale{1.0}
\end{figure}

   In Figure \ref{fig:comparison} we compare the shear rates found by calculating
the effects of each viscous mechanism individually.
In addition, we plot
the critical shear required for the Tayler-Spruit dynamo to be activated,
$\sigma_{\rm TS,crit}$ (eq. [\ref{eq:sigmatscrit}]). All models use an isothermal temperature
of $T_i=10^8\ {\rm K}$ and a WD mass of $1.37\ M_\odot$ (i.e., near the
critical mass for carbon ignition).
We calculate $\sigma_{\rm KH}$ using the approximation $\sigma_{\rm KH}\approx2N$.
Both $\sigma_{\rm BC}$ and $\sigma_{\rm TS}$ are calculated by substituting
their associated viscosities into equation (\ref{eq:angularmomentum}) and
solving for the shear using $\Omega=0.1\wk=0.67\ {\rm s^{-1}}$
and $\dot{M}=10^{-7}\ M_\odot\ {\rm yr^{-1}}$.

   Though the structure of an accreting WD is not exactly isothermal
due to the competition of compression and cooling
\citep[for example,][]{nom82}, we can still make many conclusions using
our simple model. Both $\sigma_{\rm KH}$ and $\sigma_{\rm BC}$ are
consistent with what we estimated analytically. In fact, $\sigma_{\rm KH}$
has a value within a factor of a few of what \citet{yl04} found for detailed
accreting models (see their Fig. 7). Our calculations demonstrate that
the baroclinic instability does not allow the shear to grow sufficiently for
the Kelvin-Helmholtz instability to ever be important.
Since $\sigma_{\rm BC}\ll\Omega$, the WD should exhibit nearly solid
body rotation. Also, note that this is for $\Omega=0.1\wk$. 
Larger spins result in even smaller shear since
$\sigma_{\rm BC}\propto\Omega^{-3/4}$ (eq. [\ref{eq:sigmabc}]). This is
because at higher $\Omega$ the relative amount of specific angular
momentum in the disk versus the WD surface is smaller.

\begin{figure}
\epsscale{1.2} 
\plotone{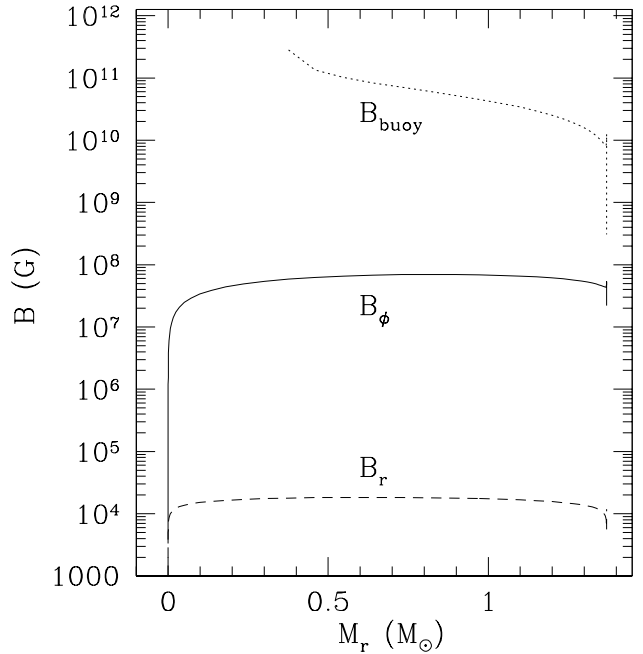}
\caption{Estimates of $B_\phi$ ({\it solid line}) and $B_r$ ({\it dashed line})
as implied by the Tayler-Spruit dynamo using the same WD model from Fig.
\ref{fig:comparison}. The fields associated with other accretion rates and
spins can be estimated using $B_\phi\propto\dot{M}^{1/2}\Omega^{-1/8}$
and $B_r\propto\dot{M}^{1/2}\Omega^{1/8}$ \citep{pb07a}. Plotted for
$H/r<1/2$ is $B_{\rm buoy}$ ({\it dotted line}), which is the critical field
needed for the buoyancy instability to occur (eq. [\ref{eq:bbuoy}]).}
\label{fig:bfields}
\epsscale{1.0}
\end{figure}

   We also consider the Tayler-Spruit dynamo in Figure
\ref{fig:comparison}. The thermal diffusivity is given by
$K =16\sigma_{\rm SB}T^3/(3c_p\kappa\rho^2)$,
where $\sigma_{\rm SB}$ is the Stefan-Boltzmann constant, $c_p$ is the specific heat
capacity at constant pressure, and $\kappa$ is the opacity.
For the opacity we include electron-scattering \citep{pac83},
free-free, and conductive contributions \citep{sch99}.
The magnetic diffusivity is set as $\eta=\pi k_{\rm B}^2Tc^2/(12 e^2 K_c)$, where
$K_c$ is the conductivity from \citet{sch99}.
The prescriptions given by \citet{spr02} imply
an even smaller shear rate than the baroclinic instability. Associated
with this small shear are steady-state radial and azimuthal
magnetic fields, which we plot in Figure \ref{fig:bfields}. This confirms
that $B_r/B_\phi\ll1$, which is expected because the toroidal field growth
is driven by shearing. The convection associated with the
simmering phase (as studied in \S \ref{sec:convection})
may destroy this magnetic field for $M_r\lesssim1\ M_\odot$,
but the field within the convectively stable region will remain.
 Whether or not these fields
are important for the later flame propagation when the WD is
incinerated or for observations of the SNe Ia is an interesting
question. According to \citet{pb07a} both $B_r,B\phi\propto\dot{M}^{1/2}$ for
the Tayler-Spruit dynamo.
Thus any process or observational diagnostic that is sensitive to
the magnetic field strength would reveal something about $\dot{M}$,
an important discriminant between progenitor models.

   There is uncertainty in applying the Tayler-Spruit formulae to
the case of an accreting WD. In the analysis presented by \citet{spr02},
it is assumed that $N>\Omega>\omega_{\rm A}$, where
$\omega_{\rm A}=B/[(4\pi\rho)^{1/2}r]$ is the Alfv\'{e}n frequency.
Such inequalities are appropriate for the radiative interior of the sun
(for which this work was originally motivated). In the WD interior
it is possible that $\Omega\gtrsim N$ since the
Brunt-V\"{a}is\"{a}l\"{a} frequency is decreased by
degeneracy effects (eq. [\ref{eq:brunt}]). \citet{dp07} consider the effects of a large spin
and find the effective viscosity of the dynamo is significantly reduced
by a factor of $(K/r^2N)^{1/6}(\Omega/N)^{1/6}(\sigma/\Omega)^{2/3}\ll1$. We hesitate
from implementing their prescriptions because
their results are based on purely heuristic arguments without
the rigorous analysis of an appropriate dispersion relation
\citep[as was provided in][]{spr06}. Since
the baroclinic instability still contributes a large viscosity, our conclusion
of solid body rotation is unchanged.

   An interesting possibility is that in the limit of large spin, a different
instability other than Tayler instability is responsible for turbulently
creating poloidal magnetic field components as is necessary for
closing the dynamo loop. The magnetorotational instability
\citep{vel59,cha60,fri69,ach78,bh91,bh92,bal95} cannot provide closure
to the dynamo since it requires $d\Omega/dr<0$, which is
opposite to what is found in the WD interior. For similar
reasons magnetic shear inabilities are also ruled out \citep{ach78}.
Since the magnetic fields plotted in Figure \ref{fig:bfields} decrease with
radius near the outer parts of the WD, it is possible that the magnetic
buoyancy instability occurs. Using the results from \citet{ach78}, \citet{spr99}
shows in the limit of $\sigma/\Omega\ll1$ that such an instability
arises for $H/r<1/2$ when
\be
	\frac{\eta}{K}\frac{N^2}{\omega_{\rm A}^2}
	+\lp\frac{r}{H}-2 \rp \frac{d\ln B}{d\ln r}\lesssim0,
\ee
where we must remember that $d\ln B/d\ln r<0$. This limit is easiest
to interpret if we solve for the magnetic field strength needed to
satisfy this inequality, which is
\be
	B\gtrsim B_{\rm buoy}
	\equiv  \frac{\lp4\pi\rho\rp^{1/2}\lp\eta/K\rp^{1/2}rN}{\lp r/H-2\rp^{1/2}\left| d\ln B/d\ln r\right|}.
	\label{eq:bbuoy}
\ee
This illustrates that the magnetic field must be sufficiently strong
in comparison to the stratification (represented by $N$) for
the buoyancy instability to occur. We plot $B_{\rm buoy}$ in
Figure \ref{fig:bfields}. Since $B_{\rm buoy}\gg B_\phi$, we conclude
that the Tayler instability limits magnetic field growth and not the buoyancy
instability. Apparently even when
$\Omega>N$ Tayler instability is the correct magnetohydrodynamic
instability for closing the shear-driven dynamo in the core of accreting WDs.

\subsubsection{Viscous Timescales and Heating}
\label{sec:timescale}

\begin{figure}
\epsscale{1.2} 
\plotone{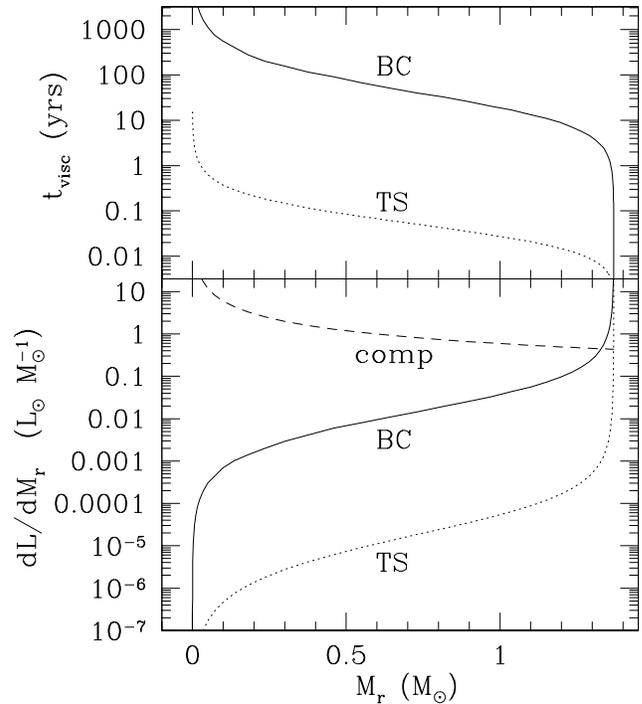}
\caption{In the top panel we compare the timescales for viscous transport of
angular momentum via the baroclinic instability ({\it solid line}) and the
Tayler-Spruit dynamo ({\it dotted line}). The bottom panel shows the
rate of energy generation for both viscous mechanisms, as well as the heating
expected from the compression of the WD from accretion
(eq. [\ref{eq:compression}] , {\it dashed line}). The WD models are the same as described in Fig.
\ref{fig:comparison}.}
\label{fig:time}
\epsscale{1.0}
\end{figure}

   These WD models provide the local viscous timescale for
angular momentum transport, $\tvisc={\rm min}[H^2,R^2]/\nu$, and
the heating provided by viscous dissipation. In the top panel of
Figure \ref{fig:time} we plot $\tvisc$ for the baroclinic instability
and the Tayler-Spruit dynamo. Both are significantly smaller than the
accretion timescale, thus our approximations presented at the beginning
of \S \ref{sec:accretion} are justified. Compositional discontinuities present a possible barrier
to angular momentum transport which we have ignored since we are focused on
the WD core. \citet{pb07a} show in the case of accreting neutron stars
that such compositional changes can inhibit turbulent mixing, but
do not alter angular momentum transport any more than introducing a slight
spin discontinuity. Our assumption of steady-state transport in the core is therefore
not affected. Figure \ref{fig:time} shows that there is a clear hierarchy of timescales. If both
mechanisms were acting, the Tayler-Spruit dynamo acts sufficiently
rapid that it limits the shear before the baroclinic instability becomes
important.

   The viscous heating per unit mass is
\be
	\epsilon_{\rm visc} = \frac{1}{2}\nu\sigma^2,
\ee
so that the total luminosity is
\be
	L_{\rm visc} = \int_0^M\epsilon_{\rm visc}dM_r.
\ee
In the bottom panel of Figure \ref{fig:time} we plot the integrand of
this expression, $dL_{\rm visc}/dM_r=\epsilon_{\rm visc}$, for the
baroclinic and Tayler-Spruit cases in units of $L_\odot\ M_\odot^{-1}$.
We denote the heating in this manner to emphasize that with these
units one can easily ``integrate by eye'' to determine the total
luminosity over any region of the WD. The process of
accretion and compression leads to additional
heating, which has been studied analytically in Appendix B of
\citet{tb04}. Using their
results, we approximate that within the degenerate core,
\be
	\frac{dL_{\rm comp}}{dM_r}
		\approx\frac{3}{5}\frac{k_{\rm B}T}{\mu_i m_p}
		\frac{\dot{M}}{M_r},
	\label{eq:compression}
\ee
where $\mu_i$ is the mean molecular weight per ion.
This is also plotted in the bottom panel of Figure \ref{fig:time}
({\it dashed line}), which shows that the viscous heating is dwarfed
by this compressional heating. Not plotted on Figure \ref{fig:time}
is the compressional heating in the non-degenerate envelope.
This total integrated contribution to the heating is
larger than that of the degenerate core for the accretion rates of
interest \citep{nom82}.

   We note that Figure \ref{fig:time} shows a model
with $\dot{M}=10^{-7}\ M_\odot\ {\rm yr^{-1}}$, and that at shallow
depths baroclinic heating is merely a factor of a few less
than compressional heating. Using the scalings derived
in \S \ref{sec:bc} we find that $dL_{\rm BC}/dM_r\propto\dot{M}^{5/4}$,
whereas $dL_{\rm comp}/dM_r\propto\dot{M}$.
Thus viscous heating from baroclinic instabilities is
important when $\dot{M}\gtrsim10^{-5}\ M_\odot\ {\rm yr^{-1}}$, which
was demonstrated in some of the models considered by
\citet{sn04}.


\section{The Convection Phase and Redistribution of Angular Momentum}
\label{sec:convection}

\begin{figure}
\epsscale{1.2} 
\plotone{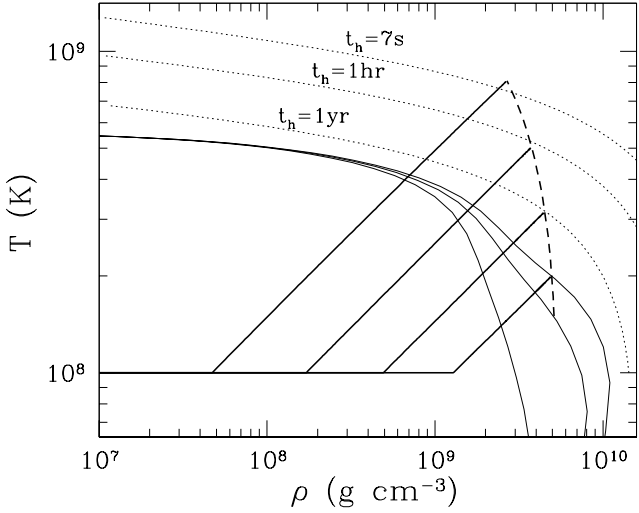}
\caption{Example thermal profiles for a $1.37\ M_\odot$ WD during
the simmering phase. Unstable carbon ignition occurs within the range denoted
by the thin solid lines \citep{yak06}. The thick dashed curve traces the
central density and temperature, ignoring compositional changes from the
burning. The thick solid lines show example thermal profiles for the WD
during different snapshots of the simmering. The non-convective region is
isothermal with $T_i=10^8\ {\rm K}$.
The dotted lines show characteristic heating timescales
as labeled.}
\label{fig:simmering}
\epsscale{1.0}
\end{figure}
   The explosive burning that incinerates the
WD in the SN Ia does not begin right at carbon ignition. Instead, the carbon simmers
for $\sim1000\ {\rm yrs}$, which we summarize in Figure \ref{fig:simmering}
\citep[also see][]{woo04,ww04,les06}. As this stage is important for setting the initial
conditions for explosive burning, there has recently been significant progress in
modeling it numerically \citep{hs02,kuh06,sw06,alm06a,alm06b}.
For our purposes here we present simpler
models that nonetheless capture many of the crucial features of this stage.
For simplicity we ignore the convective Urca process
\citep{pac72,bru73,ca75,ibe78a,ibe78b,ibe82,bw90,moc96,ste99,bk01,les05,sw06}.
Ignition first occurs when heating from carbon burning beats neutrino cooling. This
is plotted as thin solid lines that show the range of theoretical ignition
curves expected for a uniform composition of $^{12}$C and $^{16}$O,
with the middle line giving the optimal result \citep{yak06}.

   Simmering is followed by calculating a series
of hydrostatic WD models each with a different central temperature,
but with a fixed mass of $1.37\ M_\odot$. We use the same microphysics and composition
as in \S \ref{sec:comparison} and
ignore the compositional changes due to the burning.
The thick dashed line on the right of Figure \ref{fig:simmering}
traces out the central temperature as a function of
density as the simmering takes place. The density decreases
as the WD is heated and expands. The thick solid lines show example profiles demonstrating
how the convection grows as the central temperature increases.
Since the thermal conduction timescale is so long ($\sim10^6\ {\rm yrs}$), the convection
is very efficient and the
convective profile is nearly an adiabat. The adiabatic index depends sensitively
on the fact that the ions are in a highly correlated liquid state \citep{tb04},
\be
	\lp \frac{d\ln T}{d\ln \rho}\rp_{\rm ad}\approx
		\frac{0.91+0.14\Gamma^{1/3}}{1.22+0.41\Gamma^{1/3}},
\ee
where $\Gamma$ is the parameter defined in \S \ref{sec:comparison}.
For the range of models we consider, we find $(d\ln T/d\ln \rho)_{\rm ad}\approx0.49-0.52$
\citep[roughly consistent with][]{woo04}
instead of the usual $\approx2/3$ that would be appropriate if the ions were
an ideal gas. The top of the convective region is set by when it reaches
the isothermal temperature of $T_i=10^8\ {\rm K}$, in other words, where the it's entropy
matches that of the initial thermal profile \citep{hs02}.

   Also plotted in Figure \ref{fig:simmering} are dotted lines that show characteristic values
for the heating timescale, $t_h=c_pT/\epsilon$, where $\epsilon$ is the energy generation
rate for carbon burning from \citet{cf88} with strong screening included 
according to \citet{svh69}. This gives a rough estimate of the time until the burning wave
commences. For the majority of the simmering stage, the true timescale to heat
the convective region is significantly longer because the heat capacity of the entire
convective zone much be taken into account \citep[see the discussion in][]{pc08}.
The deflagration begins once $t_h\sim t_{\rm conv}$, where
\be
	t_{\rm conv}={\rm min}[H,R_{\rm conv}]/V_{\rm conv},
	\label{eq:tconv}
\ee
is the convective overturn timescale, $V_{\rm conv}$ is the characteristic convective
velocity, and we have used a mixing length equal to the minimum of either the pressure
scale height, $H$, or the radius of the convective core, $R_{\rm conv}$.
The {\it exact} condition required for the deflagration to begin depends on
what fraction of the convective zone is still responding to the energy input
and on the extreme temperature sensitivity of carbon burning
\citep[$\epsilon\propto T^{23}$,][]{woo04}. For this reason
\citet{les06} explore $t_{\rm conv}=\alpha t_h$, where $\alpha~\lesssim~1$
parameterizes this uncertainty. Since we are only roughly concerned with
resolving the end of the simmering phase, we take $t_{\rm conv}\approx t_h$
to find that simmering should end when $t_h\approx7\ {\rm s}$ at a central temperature and
density of $T_c\approx7.8\times10^8\ {\rm K}$
and $\rho_c\approx2.6\times10^9\ {\rm g\ cm^{-3}}$. This is roughly in agreement with
the estimates presented by \citet{woo04} using the Kepler stellar evolution
code \citep{wea78}.

   The series of events that take place during the simmering phase are
expected to occur regardless of the
progenitor scenario under consideration as long as ignition occurs near
the WD center. Therefore, it is important
to study any processes that may
influence the subsequent explosive burning of the SN Ia.
In the following sections we describe some of the main characteristics
of the convective region, with a focus on its mass and the velocities of
convective eddies. Afterwards we estimate how angular momentum
may be redistributed by the convection.

\subsection{Properties of the Simmering Core}

\subsubsection{Convective Core Mass}

   The boundary between the convective and non-convective zones reaches
out to densities as shallow as $\sim10^7-10^8\ {\rm g\ cm^{-3}}$
(Fig. \ref{fig:simmering})
and encompasses a mass of $M_{\rm conv}\sim0.9-1.2\ M_\odot$.
Even though the exact transitional depth depends sensitively on the
WD's isothermal temperature as well as the composition and microphysics
near the boundary \citep{pc08}, it is useful to estimate $M_{\rm conv}$.

   In order to gain some intuition for the expected dependencies we solved
for $M_{\rm conv}$ using the Lane-Emden equation for a polytrope of
index 3 in the limit of small radius. This showed us that $M_{\rm conv}$
depends on only two parameters: the mean molecular weight, $\mu_e$,
and the ratio of the isothermal and central temperatures, $T_i/T_c$.

\begin{figure}
\epsscale{1.2} 
\plotone{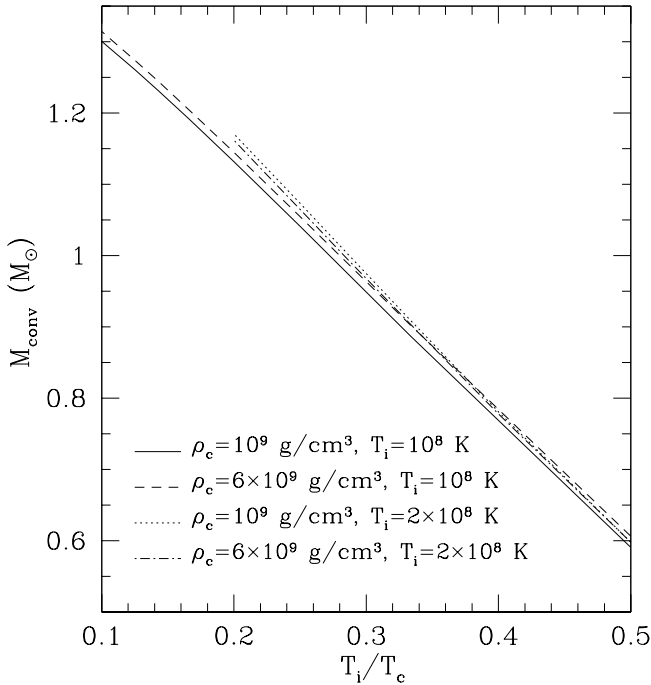}
\caption{The mass of the convective core, $M_{\rm conv}$, as a function of the ratio of
the isothermal and central temperatures, $T_i/T_c$.
We consider models with a range of central densities, $\rho_c$, and
isothermal temperatures, $T_i$, as labeled within the plot. All models
trace out a similar relation, which demonstrates that $M_{\rm conv}=M_{\rm conv}(T_i/T_c)$
is a good approximation.}
\label{fig:mconv}
\epsscale{1.0}
\end{figure}
   Unfortunately, this analytic result cannot be directly applied to the problem
at hand to estimate $M_{\rm conv}$ (because the small radius limit is not
accurate). We therefore search for the true dependence
numerically. In Figure \ref{fig:mconv} we plot $M_{\rm conv}$
versus $T_i/T_c$ for four different models that sample a range
of central densities and isothermal temperatures. This shows that the relation
is indeed very tight. A fit to the solid line in Figure \ref{fig:mconv} gives
\be
	M_{\rm conv}=1.48\ M_\odot\lp \frac{2}{\mu_e}\rp^2
		\lp 1-1.20\frac{T_i}{T_c}\rp.
	\label{eq:mconv}
\ee
This result could also be written as a ratio of
densities since the convective zone follows an adiabat with approximately
$\rho\propto T^2$.
Changes in $\mu_e$ are primarily due to $^{22}$Ne,
which gives a range $\mu_e\approx2.00-2.01$ for a
mass fractions range $X(^{22}$Ne$)=0.00-0.06$.
Therefore $\mu_e$ can change $M_{\rm conv}$ by less than 1\%. We note that
an important complication we have omitted is the presence of gradients in the C/O abundances
\citep[e.g.][]{les06}.

   This result (eq. [\ref{eq:mconv}]) is maybe
not too surprising. It is well known that for a relativistic, degenerate equation of state
that the characteristic mass of a self-gravitating object (the Chandrasekhar mass)
is independent of the central density and only depends on
$\mu_e$. In addition, nuclear reactions during the simmering may also
complicate things, which we have ignored. Nevertheless, it is an important relation to keep in mind.
It tells us that independent of the complicated previous history of a SN Ia
progenitor during the simmering phase, at any given
time $M_{\rm conv}$ depends most strongly on just a single dimensionless number.

\subsubsection{Convective Velocities}

   Besides the size of the convective zone, another important property
is the speed of the convective eddies.
Using estimates from mixing-length theory, the characteristic convective velocity,
$V_{\rm conv}$, is related to $F_{\rm conv}$ for efficient convection
via \citep{hk94}
\be
	F_{\rm conv} = \frac{c_pT}{\mathcal{Q}gH}\rho V_{\rm conv}^3
	\approx \rho V_{\rm conv}^3,
	\label{eq:vconv}
\ee
where we have used a mixing-length that is equal to the scale height, $H$.
If the convective region was static during simmering, then $F_{\rm conv}$
would simply be equal to the integrated nuclear energy generation. Instead
the convection zone is growing and energy must be expended to heat new
material. Also, the convective flux must be nearly zero at the convective boundary
because of the long conductive timescale here ($\sim10^6\ {\rm yrs}$).
These effects cause a decreased $F_{\rm conv}$ in the outer
parts of the convective region, which we investigate in more detail in a separate
study \citep{pc08}. We borrow these results for presenting $V_{\rm conv}$ here.

\begin{figure}
\epsscale{1.2} 
\plotone{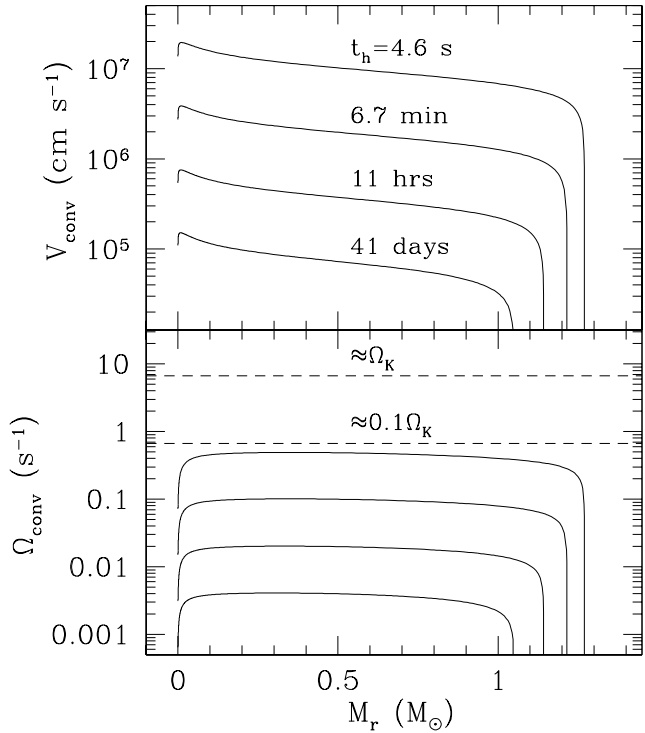}
\caption{In the top panel we plot the characteristic convective velocity,
$V_{\rm conv}$, as a function of mass coordinate for various times during
the simmering phase. Each line is labeled by the heating timescale,
$t_h=c_pT/\epsilon$, which is meant as a rough estimate for the time until the
deflagration begins. In the bottom panel
we plot the convective eddy turnover frequency, $\Omega_{\rm conv}=t_{\rm conv}^{-1}$
(eq. [\ref{eq:tconv}]), for the same four models. The dashed lines denote a reasonable
range for the WD spin $\Omega\approx(0.1-1.0)\Omega_{\rm K}$.}
\label{fig:vconv}
\epsscale{1.0}
\end{figure}

  In the upper panel of Figure \ref{fig:vconv} we plot $V_{\rm conv}$ as a function
of mass coordinate using equation (\ref{eq:vconv}). The characteristic value of
$V_{\rm conv}\sim10^7\ {\rm cm\ s^{-1}}$ during the late stages of convection
is consistent with the analytic estimates of \citet{woo04}.
The turbulent conditions prepared in the core will be important for understanding
the subsequent propagation of bubbles and burning. For example,
\citet{zd07} show that the turbulence may act as a viscous drag
with the largest effect on the smallest bubbles. Another interesting feature is
that the material above the convection is devoid of this turbulence.
The burning properties could change significantly as a flame passes into this
relatively ``quiet'' region.

   The importance of the WD spin for the convection is determined
by the convective Rossby number
\be
	Ro = \Omega_{\rm conv}/\Omega,
\ee
where $\Omega_{\rm conv}=t_{\rm conv}^{-1}$ (see eq. [\ref{eq:tconv}]) is the eddy turnover
frequency. In the bottom panel of Figure \ref{fig:vconv} we plot $\Omega_{\rm conv}$
along with the expected range of spins relevant for
SN Ia progenitors ({\it dotted lines}). For the entirety of the simmering
stage we find $Ro\lesssim1$, showing that spin should have a non-negligible
influence on the convection \citep[for example, see][]{kuh06}. This will
be an important consideration in the following sections.

\subsection{Rotation Profiles during Simmering}

   We next consider how the spin changes as convection grows in the core.
For the $10^5-10^9\ {\rm yrs}$ that
accretion takes place we have shown that
hydrodynamic or magnetohydrodynamic instabilities severely limit
shearing. During the last stages of simmering the WD
is changing on the timescale of hours to minutes, so these viscosities will
have less influence and stronger shearing is possible.

   Given the conclusions made in \S \ref{sec:accretion}, we consider
uniform rotation to be a reasonable starting point for when
carbon first ignites. Heating and expansion generally causes the
WD to spin down (like the freshman physics figure skater problem). In the following
calculations we assume that angular momentum is conserved locally on spherical
mass shells in non-convective regions. The mixing by convective eddies is
important  for determining the spin within the convective core.
Such effects are difficult to estimate, so we
consider two possible rotation laws.

\subsubsection{Uniform Spin within the Convection}

   The first rotation law we explore is that convection enforces solid
body rotation. This is the simplest and most na\"{i}ve assumption one can
use. The study by \citet{kum95} argues that such a rotation law is reasonable
when convective eddies scatter elastically. Unfortunately, it is
not clear whether such a picture of the eddies as separate fluid elements is
appropriate. Elastic scattering is in fact counter to the standard
assumption in mixing-length theory that eddies deposit their entire
angular momentum and entropy into the ambient fluid with each scattering.
This result also assumes an isotropic scattering whereas rotation clearly
breaks this symmetry. In the anisotropic case \citet{kum95} find that
in principle angular momentum can be transferred outward by convection,
an idea we consider in more detail in the next section.

\begin{figure}
\epsscale{1.2} 
\plotone{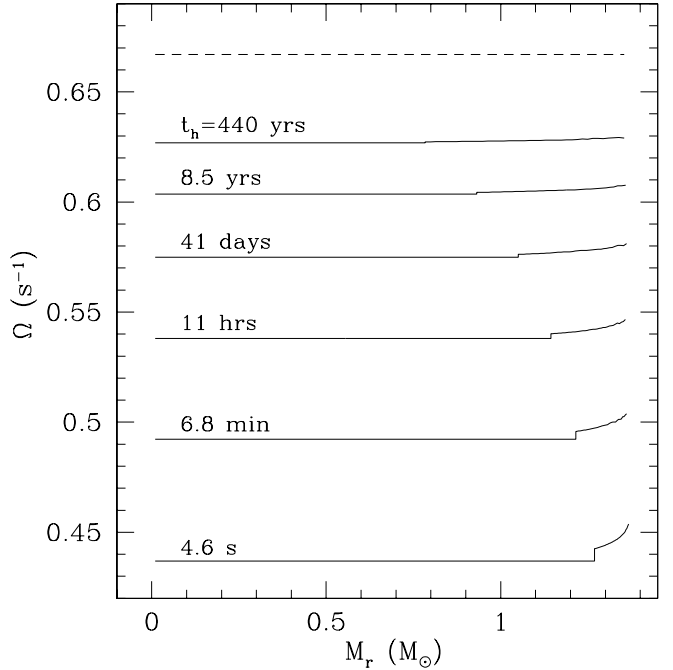}
\caption{The rotation profile versus mass coordinate for various times during the simmering
stage assuming that uniform rotation is enforced within the convective
zone. Each curve is labeled by its associated heating timescale, $t_h$.
The initial spin is taken to be $\Omega=0.1\Omega_{\rm K}=0.67\ {\rm s^{-1}}$
({\it dashed line}) with a mass of $1.37\ M_\odot$.}
\label{fig:shear_solid}
\epsscale{1.0}
\end{figure}

   In Figure \ref{fig:shear_solid} we plot the evolution of the WD spin
during the simmering stage assuming solid body rotation within the convective
zone. The initial spin is taken to be uniform with $\Omega=0.67\ {\rm s^{-1}}$
({\it dashed line}).
The overall trend is spin-down from heating and expansion.
Uniform rotation in the convective zone leads to shear at its
top with a discontinuous velocity of $\Delta V\sim10^5-10^6\ {\rm cm\ s^{-1}}$
at late times.

\subsubsection{Outward Transport of Angular Momentum by Convection}

   The other rotation law that we consider is motivated by numerical
simulations and observations of the Sun. In theoretical and numerical studies
where $Ro\lesssim1$, a generic feature is that angular momentum is transported
outward away from the poles and towards the equator \citep{gil79,mie00,bru02,bro04}.
This appears to be controlled by the largest
scale eddies that are most influenced by rotation \citep{bro04}. Such features are qualitatively
consistent with helioseismic measurements that map the outer convective region
of the Sun \citep[][and references therein]{tho03}. The main disparity is that theory
and numerics generally result in a more Taylor-Proudman like spin profile whereas
the Sun's velocity contours are more radial. This may be due to the tachocline
\citep{mie06}, which is present in cases with surface convection. The study of
core convection in rotating A-stars by \citet{bro04}, which also finds
a Taylor-Proudman spin profile, may be the most relevant comparison
to the WD case.

   To mimic the general features of the spin profiles described above we
consider the following rotation law,
\be
	\Omega(r,\theta)
		= \left[ \beta\lp \frac{r\sin\theta}{R_{\rm conv}}\rp+1\right]\Omega_c,
	\label{eq:rotationlaw}
\ee
where $\Omega_c$ is the spin at the WD's center, $\beta$ is the fractional change in spin
across the convective zone, and $\theta$ is the latitude measured from the pole.
For calculational purposes, we choose $\beta=0.4$.
This is in reasonable agreement with the results of \citet{bro04} and solar
observations \citep{tho03}, which both have a similar value of $Ro\sim0.1$ in comparison
to our case here.

\begin{figure}
\epsscale{1.2} 
\plotone{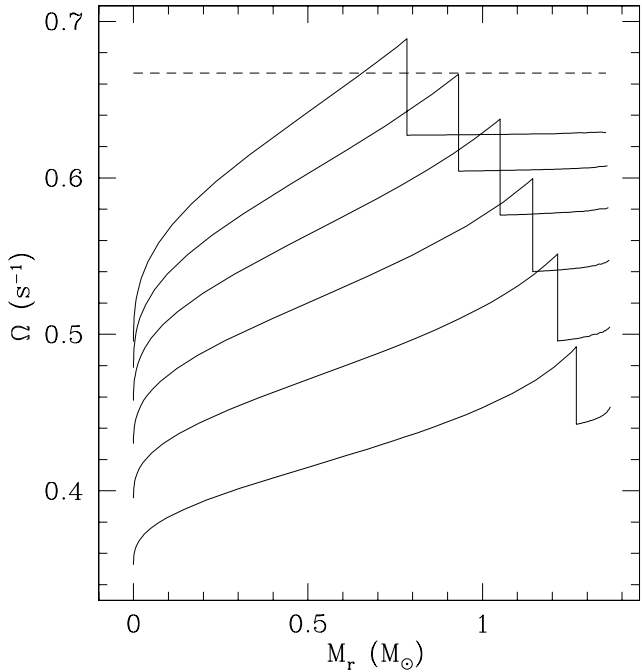}
\caption{The spin {\it along the equator} versus mass coordinate assuming a rotation law
within the convective zone give be eq. (\ref{eq:rotationlaw}),
using the same background models as Fig. \ref{fig:shear_solid}.}
\label{fig:shear_incr}
\epsscale{1.0}
\end{figure}
   In Figure \ref{fig:shear_incr} we plot the spin profile along the equator
using equation (\ref{eq:rotationlaw}) as the rotation law within the convection.
The outward transport of angular momentum is much more prominent
in comparison to the uniform rotation case considered earlier.
Since the rotation is cylindrical, at higher latitudes the
shear is significantly smaller. The typical velocity jump at the equator
is $\Delta V\sim10^6-10^7\ {\rm cm\ s^{-1}}$, which may be comparable
to the speeds of the burning fluid elements that will be buoyantly rising
through the WD once explosive carbon ignition occurs. This could in
principle shear out the burning, enhancing it because of the increased
surface area.

\subsubsection{Heating at the Convective Boundary}

   The shear present at the convective/non-convective boundary
represents free energy that has been made available because of angular momentum
transport by convection. If this shear persists throughout the convective
stage, then it may have interesting effects on the subsequent flame
propagation. The shear is not smeared out by convective overshooting. Using an
overshooting length of
$L\approx V_{\rm conv}^2/(2g)$ and a sound speed $c_s$,
we estimate that $L/H\sim (V_{\rm conv}/c_s)^2\ll1$,
since the convective velocities are very sub-sonic.
On the other hand, viscous processes may act at this
interface, damping out the shear and leading to heating. Assuming all
of the shear is converted to heat gives a local heating per unit mass of
$\sim(\Delta V)^2$. This is basically an upper limit of the energy
generation rate at any given time.

\begin{figure}
\epsscale{1.2} 
\plotone{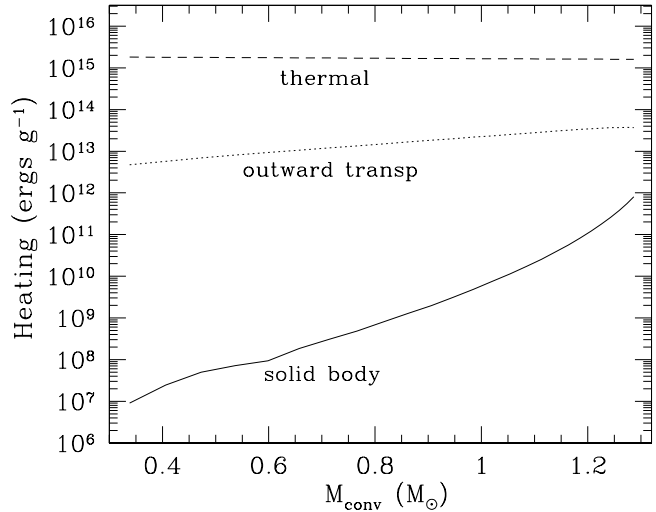}
\caption{The heating available from the shear at the convective/non-convective
boundary as a function of the boundary's mass coordinate. These are found
for both of the rotation laws we consider for the convection of
solid-body rotation ({\it solid line})
and outward angular momentum transport ({\it dotted line}). As a comparison
we show the thermal energy at the top of the convection zone, $c_pT$
({\it dashed line}).}
\label{fig:deltav}
\epsscale{1.0}
\end{figure}

   In Figure \ref{fig:deltav} we plot the free energy per unit mass available from
the shear at the convective/non-convective boundary for both rotation
laws. This is shown to be less than the thermal energy
at the same location, $c_pT$. We note that these results are for an initial spin of
$\approx0.1\Omega_{\rm K}$ (chosen so that the spin would have minimal impact
on hydrostatic balance). Since $\Delta V\propto\Omega$,
the heating can be up to $\approx100$ times larger than what we estimate here. In the
case where angular momentum is transported outward by convection, dissipation
of shear may be able to heat the outer layers of the star. This may be an important
effect for priming the thermal profile and expanding the WD before the
flame propagates past.
   

\section{Discussion and Conclusion}
\label{sec:theend}

   In this study we have investigated the internal rotation profile for
SN Ia progenitors, focusing on the stages of accretion and
simmering. Since these progenitors are formed exclusively in binary systems, they contain
a enormous amount of angular momentum. This is important for
the WD structure and the subsequent flame propagation during the explosive burning of a SN Ia.

   During the accretion phase we compared three different mechanisms for
angular momentum transport. Either the baroclinic instability or the Tayler-Spruit
dynamo limit the shear to values so low that the Kelvin-Helmholtz instability
cannot occur. We argue that the WD will have nearly uniform rotation,
consistent with the results of \citet{sn04}.

  This conclusion has important implications for
extremely energetic SNe Ia such as SN2003fg \citep[SNLS-03D3bb,][]{how06},
which has been argued to be from a super-Chandrasekhar WD \citep{jef06}.
One well-known way to get a mass this high is via strong differential rotation
\citep[][and references therein]{st83}. If uniform rotation is required then
the WD mass can only be increased by a few percent over the normal Chandrasekhar
limit. We should also mention that just because the WD
mass is higher, the chemical yields of the deflagration or detonation of a strongly
differentially rotating WD
will not necessarily match the high nickel mass needed for SN2003fg
\citep{ste92,pfa06}. For this reason, other interpretations have been
suggested for this SN Ia \citep{hil07}.

   The simmering phase is another opportunity for differential rotation to
develop within the WD. This occurs due to heating and expansion as well
as angular momentum mixing by convection. The full impact of the convection
is hard to determine with certainty. Given the characteristic convective
velocities, we expect $Ro\lesssim1$, thus the influence of rotation cannot
be discounted. We consider two rotation laws for the convective zone, a
uniform rotation case and another case where the spin increases outward.
In either case the most dramatic shearing appears at the convective/non-convective
boundary. This shearing may have an impact on the propagation of flames or
bubbles, and in some extreme circumstances may hold enough free energy
to alter the thermal profile of the non-convective envelope.

   Given the impact rotation could potentially have for SNe Ia,
the time appears ripe for including such effects in simmering and
flame propagation calculations. Simulations such as those performed
by \citet{bro04} are not yet able to reach the characteristic fluid parameters
expected during the simmering stage
\citep[such as the Reynolds and Rayleigh numbers; see the summary in][]{woo04,ww04}.
Nevertheless, $Ro\approx0.1$ is well within reach, as has been shown by the simulations
of \citet{kuh06}. The biggest difficulty for the future will be following the secular evolution
of the WD as the convective zone grows. Since the carbon burning is so temperature
sensitive, the majority of the evolution takes place at late times
(see Fig. \ref{fig:shear_solid} and \ref{fig:shear_incr}). This will help limit
the duration that must be simulated.

\acknowledgements
I thank Lars Bildsten for comments on a previous draft and
Dmitry Yakovlev for sharing carbon ignition
curves. I am also grateful to Matt Browning, Phil Chang, and Eliot Quataert for helpful
discussions. This work was partially supported by the National Science Foundation
under grants PHY 99-07949 and AST 02-05956.



\begin{thebibliography}

\bibitem[Acheson(1978)]{ach78}
Acheson D.J., 1978, Phil. Trans. Roy. Soc. Lond. A, 289, 459



\bibitem[Almgren et al.(2006a)]{alm06a}
Almgren, A. S., Bell, J. B., Rendleman, C. A., \& Zingale, M. 2006a, ApJ, 637, 
922

\bibitem[Almgren et al.(2006b)]{alm06b}
Almgren, A. S., Bell, J. B., Rendleman, C. A., \& Zingale, M. 2006b, ApJ, 637, 
922

\bibitem[Balbus(1995)]{bal95}
Balbus S.A., 1995, \apj, 453, 380

\bibitem[Balbus \& Hawley(1991)]{bh91}
Balbus S.A. \& Hawley J.F., 1991, \apj, 376, 214

\bibitem[Balbus \& Hawley(1992)]{bh92}
Balbus S.A. \& Hawley J.F., 1992, \apj, 400, 610

\bibitem[Barkat \& Wheeler(1990)]{bw90}
Barkat, Z., \& Wheeler, J. C. 1990, ApJ, 355, 602

\bibitem[Bisnovatyi-Kogan(2001)]{bk01}
Bisnovatyi-Kogan, G. S. 2001, MNRAS, 321, 315

\bibitem[Browning et al.(2004)]{bro04}
Browning, M. K., Brun, A. S., \& Toomre, J. 2004, \apj, 601, 512

\bibitem[Bruenn(1973)]{bru73}
Bruenn S. 1973, ApJ, 183, L125

\bibitem[Brun \& Toomre(2002)]{bru02}
Brun, A. S. \& Toomre, J. 2002, \apj, 570, 865

\bibitem[Caughlan \& Fowler(1988)]{cf88}
Caughlan, G. R., \& Fowler, W. A. 1988, At. Data Nucl. Data Tables, 40, 283

\bibitem[Chabrier \& Potekhin(1998)]{cp98}
Chabrier, G., \& Potekhin, A. Y. 1998, Phys. Rev. E, 58, 4941

\bibitem[Chamulak et al.(2007)]{cha07}
Chamulak, D. A., Brown, E. F., Timmes, F. X., \& Dupczak, K. 2007,
submitted to \apj

\bibitem[Chandrasekhar(1960)]{cha60}
Chandrasekhar S., 1960, Proc. Nat. Acad. Sci., 46, 253

\bibitem[Couch \& Arnett(1975)]{ca75}
Couch, R. G., \& Arnett, W. D. 1975, ApJ, 196, 791

\bibitem[Cumming \& Bildsten(2000)]{cb00}
Cumming, A. \& Bildsten, L. 2000, \apj, 544, 453

\bibitem[Denissenkov \& Pinsonneault(2007)]{dp07}
Denissenkov, P. A. \& Pinsonneault, M. 2007, \apj, 655, 1157

\bibitem[Fricke(1969)]{fri69}
Fricke K., 1969, \aap, 1, 388

\bibitem[Fujimoto(1987)]{fuj87}
Fujimoto, M. Y. 1987, \aap, 176, 53

\bibitem[Fujimoto(1988)]{fuj88}
Fujimoto, M. Y. 1988, \aap, 198, 163

\bibitem[Fujimoto(1993)]{fuj93}
Fujimoto, M. Y. 1993, \apj, 419, 768


\bibitem[Gilman(1979)]{gil79}
Gilman, P. A. 1979, \apj, 231, 284


\bibitem[Hansen \& Kawaler(1994)]{hk94}
Hansen, C. J. \& Kawaler, S. D. 1994, Steller Interiors: Physical Principles,
Structure, and Evolution (Berlin: Springer)


\bibitem[Hillebrandt \& Niemeyer(2000)]{hn00}
Hillebrandt, W. \& Niemeyer, J. C. 2000, ARA\&A, 38, 191

\bibitem[Hillebrandt et al.(2007)]{hil07}
Hillebrandt, W., Sim, S. A., \& R\"{o}pke, F. K. 2007, \aap, 465, L17

\bibitem[H\"{o}flich \& Stein(2002)]{hs02}
H\"{o}flich, P. \& Stein, J. 2002, \apj, 568, 779

\bibitem[Howell et al.(2006)]{how06}
Howell, D. A., et al. 2006, \nat, 443, 308

\bibitem[Iben(1978a)]{ibe78a}
Iben, I. 1978a, ApJ, 219, 213

\bibitem[Iben(1978b)]{ibe78b}
Iben, I. 1978b, ApJ, 226, 996

\bibitem[Iben(1982)]{ibe82}
Iben, I. 1982, ApJ, 253, 248

\bibitem[Jeffrey et al.(2006)]{jef06}
Jeffery, D. J., Branch, D., \& Baron, E. 2006, submitted to \apj\ (astro-ph/0609804)





\bibitem[Kuhlen et al.(2006)]{kuh06}
Kuhlen, M., Woosley, S. E., \& Glatzmaier, G. A. 2006, \apj, 640, 407

\bibitem[Kumar et al.(1995)]{kum95}
Kumar, P., Narayan, R., \& Loeb, A. 1995, \apj, 453, 480


\bibitem[Lesaffre et al.(2006)]{les06}
Lesaffre, P., Han, Z., Tout, C. A., Podsiadlowski, Ph., \& Martin, R. G., 2006, 
MNRAS, 368, 187 

\bibitem[Lesaffre et al.(2005)]{les05}
Lesaffre, P., Podsiadlowski, Ph., \& Martin, C. A., 2005, 
MNRAS, 356, 131

\bibitem[Livio \& Pringle(1998)]{lp98}
Livio, M., \& Pringle, J. E. 1998, \apj, 505, 339 


\bibitem[MacDonald(1979)]{mac79}
MacDonald, J. 1979, Ph. D. Thesis, Cambridge University


\bibitem[Miesch(2000)]{mie00}
Miesch, M. S. 2000, Sol. Phys., 192, 59

\bibitem[Miesch et al.(2006)]{mie06}
Miesch, M. S., Brun, A. S., \& Toomre, J. 2006, \apj, 641, 618

\bibitem[Mochkovitch(1996)]{moc96}
Mochkovitch, R. 1996, \aap, 311, 152

\bibitem[Nandkumar \& Pethick(1984)]{np84}
Nandkumar, R., \& Pethick, C. J. 1984, MNRAS, 209, 511 

\bibitem[Nomoto(1982)]{nom82}
Nomoto, K. 1982, \apj, 253, 798

\bibitem[Paczy\'{n}ski(1972)]{pac72}
Paczy\'{n}ski, B. 1972, Astrophys. Lett., 11, 53

\bibitem[Paczy\'{n}ski(1983)]{pac83}
Paczy\'{n}ski, B. 1983, \apj, 267, 315

\bibitem[Paczy\'{n}ski(1991)]{pac91}
Paczy\'{n}ski, B. 1991, \apj, 370, 597

\bibitem[Pfannes(2006)]{pfa06}
Pfannes, J. M. M. 2006, Ph. D. Thesis, Universit\"{a}t W\"{u}rzburg

\bibitem[Piro \& Bildsten(2004)]{pb04}
Piro, A. L. \& Bildsten, L. 2004, \apj, 610, 977

\bibitem[Piro \& Bildsten(2007a)]{pb07a}
Piro, A. L. \& Bildsten, L. 2007a, \apj, 663, 1252

\bibitem[Piro \& Bildsten(2007b)]{pb07b}
Piro, A. L. \& Bildsten, L. 2007b, accepted for publication in \apj\ (arXiv:0710.1600)

\bibitem[Piro \& Chang(2008)]{pc08}
Piro, A. L. \& Chang, P. 2008, accepted for publication in \apj

\bibitem[Popham \& Narayan(1991)]{pn91}
Popham, R., \& Narayan, R. 1991, \apj, 370, 604


\bibitem[Riess et al.(2004)]{rie04}
Riess, A. G., et al. 2004, \apj, 607, 66

\bibitem[Saio \& Nomoto(2004)]{sn04}
Saio, H. \& Nomoto, K. 2004, \apj, 615, 444

\bibitem[Salpeter \& van Horn(1969)]{svh69}
Salpeter, E. E., \& van Horn, H. M. 1969, \apj, 155, 183

\bibitem[Schatz et al.(1999)]{sch99}
Schatz, H., Bildsten, L., Cumming, A. \& Wiescher, M. 1999,
\apj, 524, 1014

\bibitem[Shapiro \& Teukolsky(1983)]{st83}
Shapiro, S. L., \& Teukolsky, S. A. 1983, Black Holes, White Dwarfs, and
Neutron Stars (New York: Wiley)

\bibitem[Sion(1999)]{sio99}
Sion, E. M. 1999, PASP, 111, 532

\bibitem[Spruit(1999)]{spr99}
Spruit, H. C. 1999, \aap, 349, 189

\bibitem[Spruit(2002)]{spr02}
Spruit, H. C. 2002, \aap, 381, 923

\bibitem[Spruit(2006)]{spr06}
Spruit, H. C. 2006, preprint (astro-ph/0607164)



\bibitem[Stein et al.(1999)]{ste99}
Stein, J., Barkat, Z., \& Wheeler, J. C. 1999, ApJ, 523, 381

\bibitem[Stein \& Wheeler(2006)]{sw06}
Stein, J. \& Wheeler, J. C. 2006, ApJ, 643, 1190

\bibitem[Steinmetz et al.(1992)]{ste92}
Steinmetz, M., M\"{u}ller, E., \& Hillebrandt, W. 1992, \aap, 254, 177

\bibitem[Tayler(1973)]{tay73}
Tayler, R. J. 1973, \mnras, 161, 365

\bibitem[Thompson et al.(2003)]{tho03}
Thompson, M. J., Christensen-Dalsgaard, J., \& Miesch, M. S. 2003, ARA\&A, 
41, 599

\bibitem[Townsley \& Bildsten(2004)]{tb04}
Townsley, D. M.\& Bildsten, L. 2004, \apj, 600, 390


\bibitem[Velikhov(1959)]{vel59}
Velikhov E.P., 1959, J. Exp. Theoret. Phys., 36, 1398


\bibitem[Weaver et al.(1978)]{wea78}
Weaver, T. A., Woosley, S. E., \& Zimmerman, G. B. 1978, ApJ, 225, 1021 

\bibitem[Woosley et al.(2004)]{woo04}
Woosley, S. E., Wunsch, S., \& Kuhlen, M. 2004, \apj, 607, 921

\bibitem[Wunsch \& Woosley(2004)]{ww04}
Wunsch, S. \& Woosley, S. E. 2004, \apj, 616, 1102

\bibitem[Yakovlev et al.(2006)]{yak06}
Yakovlev, D. G., Gasques, L. R., Afanasjev, A. V., Beard, M., \& Wiescher, M.
2006, Phys. Rev. C, 74, 035803

\bibitem[Yoon \& Langer(2004)]{yl04}
Yoon, S.-C. \& Langer, N. 2004, \aap, 419, 623

\bibitem[Zahn(1992)]{zah92}
Zahn, J.-P. 1992, \aap, 265, 115

\bibitem[Zingale \& Dursi(2007)]{zd07}
Zingale, M. \& Dursi, L. J. 2007, \apj, 656, 333

\end{thebibliography}
\end{document}